\documentclass[12pt]{iopart}
\usepackage{cite,amssymb,amsfonts,graphicx,bm,dcolumn,mathrsfs}
\newcommand{\bea}{\begin{eqnarray}}
\newcommand{\eea}{\end{eqnarray}}
\newcommand{\be}{\begin{equation}}
\newcommand{\ee}{\end{equation}}
\newcommand{\dd}{\mbox{d}}
\newcommand{\ii}{{\rm i}}
\usepackage{epsfig}

\begin{document}

\title[Noisy oscillators with bimodal frequency distribution]{Phase diagram of noisy systems of coupled
oscillators with a bimodal frequency distribution}

\author{Alessandro Campa}
\address{National Center for Radiation Protection and
Computational Physics, Istituto Superiore di Sanit\`{a},
Viale Regina Elena 299, 00161 Roma, Italy}
\ead{alessandro.campa@iss.it}

\begin{abstract}
We study the properties of large systems of globally coupled oscillators in the presence of noise. When the
distribution of the natural frequencies of the oscillators is bimodal and its analytical continuation in the complex plane has only few poles
in the lower half plane,
the dynamics of the system, governed by a Fokker-Planck equation for the single particle distribution function, can be reduced to a system
of ordinary differential equations describing the dynamics of suitably defined order parameters, the first ones of which are related to the
usual synchronization order parameter. We obtain the full phase diagram of the oscillator system, that shows a very rich behaviour,
with regions characterized by synchronized states, regions with periodic states, and others with bi-stability, associated to the presence of
hysteresis. The latter phenomenon is confirmed by numerical simulations ot the full system of coupled oscillators. We compare our results with
those previously obtained for noiseless systems, and we show that for increasing noise the phase diagram changes qualitatively, tending to the simple
diagram that is found for systems with unimodal frequency distributions.
\end{abstract}

\submitto{\jpa}
\maketitle

\section{Introduction}
\label{intro}

The phenomenon of collective synchronization in systems made of a large population of coupled oscillatory units is now recognized
as a very important subject of investigation, since it is naturally found in many different situations \cite{pikovsky2001}. Although the units
are characterized by different natural frequencies, they can spontaneously synchronize and oscillate at a common frequency. This cooperative
effect can be found in physical and biological systems, like flashing in unison by groups of fireflies \cite{buck1988},
voltage oscillations at a common frequency in an array of current-biased Josephson junctions \cite{wiesen1998}, synchronized firings of
cardiac pacemaker cells \cite{winfree1980}, metabolic synchrony in yeast cell suspensions \cite{bier2000}, phase synchronization in electrical
power distribution networks \cite{fila2008}, animal flocking behaviour \cite{ha2010}. A survey of the examples occurring in nature is
in \cite{strogatz2003}.

It is not possible to overestimate the great importance of the introduction of the Kuramoto model for the
theoretical study of synchronization in systems of interacting oscillators \cite{kura1,kura2}. The model makes several assumptions;
in particular, it assumes that the oscillators are represented by a single dynamical variable, the phase, and that the coupling strength $K$
is the same between all pairs of oscillators; besides, the interaction is very simple, depending only on the sine of the pahse
difference of the pair.  In spite of these simplifying characeristics, the model captures the essential
physics of the dynamics, in which the interaction can induce a macroscopic fraction of the oscillators, each one with a proper frequency
drawn from a given distribution $g(\omega)$, to spontaneously synchronize.

The original model has been extended along several directions. In the Kuramoto model the proper frequencies of the oscillators are
quenched variables. However, the recognition of the fact that the natural frequency of each oscillatory unit can fluctuate for various reasons
(we remind that the simple phase representation can model in an effective way a quite complex physical or biological unit), has led to the
introduction of noise in the dynamics, transforming the original deterministic equations in Langevin equations \cite{saka}. This noise can be thought of
as mimicking the effect of frequency fluctuations; thus the strenght of the noise is directly related to the amplitude of these fluctuations. Another
generalization has been the introduction of inertia, proposed as a way to improve the modelization of the approach to synchronization \cite{ermen}. Taken
together, the two generalizations result in a second-order dynamical system subject to noise \cite{aceb2000,gupta2014}.

A prominent role is played by the frequency distribution function $g(\omega)$. Most of the reasearch has been devoted to the case of a unimodal
distribution, i.e., to the case where $g(\omega)$ has a unique maximum at a frequency $\omega=\omega_0$, around which it is symmetrical,
decaying monotonically to zero for increasing $|\omega - \omega_0|$. In this case the Kuramoto model has a synchronization transition at the
value $K_c=\frac{2}{\pi g(\omega_0)}$ of the coupling; for smaller values the oscillators do not synchronize and each one oscillates with its own
proper frequency, while for $K>K_c$ a fraction of the oscillators synchronize at $\omega=\omega_0$, the fraction increasing continuosly
for increasing $K$, from $0$ at $K_c$ to $1$ when $K$ becomes very large (formally for $K\to \infty$) \cite{strogatz2000}. This picture holds
for the model augmented with noise, although the threshold value for $K$ depends on the noise strength \cite{saka,reviewjstat}. However,
for more general frequency distributions, the overall scenario can be more complex, as shown by, e.g., numerical simulations of the dynamics when
$g(\omega)$ is bimodal, having two equal maxima at two different frequencies \cite{kurabrief}: the system of oscillators can present bi-stability and
also seemingly periodic asymptotic states.

Independently from the role played by the distribution $g(\omega)$ in determining the possible synchronized states of the system of oscillators,
another problem, in a theoretical analysis, is represented by the study of the dynamics itself. As shown in the next Section, the dynamics can be
described, at least in the limit of a very large number of oscillators (formally, for $N\to \infty$), by a Fokker-Planck equation for the
time dependent single particle distribution function. As will be clear, the asymptotic stationary distributions at large times are related, when different from
a homogeneous distribution, to the synchronized states. However, the analysis of the full dynamics as determined by the Fokker-Planck equation is not a trivial
task, and it is much more feasible to find its possible stationary states and the behaviour of the system, perturbatively, when the coupling $K$ is near
the critical value $K_c$ \cite{strogatz2000,kurabrief}. In this respect, a breakthrough has been provided, in the study of noiseless systems, by the Ott-Antonsen
ansatz \cite{ott-anton}, that allows to reduce the dynamics to that of a single Fourier component of the single particle distribution (although the dynamical
system is still infinite dimensional, since the Fourier component depends not only on time, but also on $\omega$). This reduction goes further
if the frequency distribution function $g(\omega)$ can be analytically continued in the complex plane and this continuation has only few poles in the lower half
plane; then, it is possible to study directly the full dynamics of few variables (practically, two real equations for each pole of the
analytical continuation) that are simply related to the synchronization of the system; this will be completely clear in the next Section. One has a low-dimensional
system of ordinary differential equations.

While this approach has confirmed the picture previously described for unimodal distributions $g(\omega)$, its application to a bimodal $g(\omega)$,
with two poles in the lower half complex plane,
has revealed the great richness of the possible states of the system of oscillators \cite{martens2009}. In this latter work, the results have been obtained by the
study of a system of four ordinary differential equations, reduced to two by simple and plausible physical arguments. The analysis of a twodimensional
dynamical system has made it possible to obtain a full phase diagram of the asymptotic states. Unfortunately, the introduction of noise prevents
the use of the Ott-Antonsen ansatz, as will be shown. On the other hand, the study of a system where $g(\omega)$ has the above mentioned property
of having few poles in the lower half of the complex plane, allows to reduce the dynamics to a system of ordinary differential equations. Even if
in principle the system is (countably) infinite dimensional, it is possible to restrict the analysis to a truncated system of few tens of equations.
This is the central point of this paper. We will consider the possible synchronized states, and the periodic asymptotic states, of a noisy system of coupled
oscillators, in which the frequency distribution $g(\omega)$ is bimodal and has two poles in the lower half of the complex plane.
Although the number of equations is not equal to $2$ as for the noiseless system, preventing an almsot complete analytical study of the solutions,
the system can be very rapidly analyzed numerically, and in particular also the stability of synchronized states can be studied, something that is
extremely difficult for the full Fokker-Planck equation.

The structure of the paper is as follows. In Section \ref{derequat} we introduce the model and we derive the system of equations. In Section \ref{statstates}
we study the stationary states and their stability, while in Secion \ref{wavestates} we focus on the periodic asymptotic states; in these two sections, we concentrate
on a given values of the noise strength, and we present the full phase diagram (in the parameter space) of the system, that interestingly includes
regions of bi-stability; besides, we make a brief analysis of the
bifurcations associated to the transitions from one stationary state to another (or from a stationary state to a periodic state). In Section \ref{largerd}
we consider the phase diagram at increasingly larger noise strength, showing how the diagram simplifies, approaching the one valid for a unimodal distribution.
In Section \ref{simdyn} we compare the results of our analysis with those of a numerical simulation of the full system of noisy coupled oscillators. Besides
showing the agreement of the two evaluations, the section focuses to the presence of hysteresis in the dynamics, directly associated to the presence
of bi-stability regions in the phase diagram. Section \ref{discuss} presents a discussion and draws some conclusions.

\section{Derivation of the system of equations}
\label{derequat}

The Langevin equations describing the dynamics of $N$ oscillators that interact with a coupling as in the Kuramoto
model, are \cite{saka}:
\be
\frac{\dd \theta_i}{\dd t} = \omega_i -\frac{K}{N}\sum_{j=1}^N \sin \left( \theta_i - \theta_j \right)
+\eta_i(t) \, , \,\,\,\,\,\,\,\,\,\,\,\,\,\,\,\,\,\, i=1,\dots,N, \,
\label{langeq}
\ee
where $\theta_i \in [0,2\pi)$ is the phase of the $i$-th oscillator, $\omega_i$ is its natural frequency,
and $K$ is the coupling constant. The stochastic noise $\eta_i(t)$ is independent from those of the other
oscillators, and each $\eta_i(t)$ is Gaussian distributed at each time, while the noises at different times are
uncorrelated. Then we have the expectation values (averaging over noise realizations):
\be
\langle \eta_i(t) \rangle =0 \, , \,\,\,\,\,\,\,\,\,\,\,\,\,\,\,\,\,\,
\langle \eta_i(t) \eta_j(t') \rangle = 2 D \delta_{ij} \delta (t-t') \, ,
\label{noisestat}
\ee
where the coefficient $D$ characterizes the noise intensity. The natural frequencies $\omega_i$ are distributed
according to a given frequency distribution function $g(\omega)$. As anticipated above, we will treat the case of a symmetric
bimodal frequency distribution, given in particular by the sum of two Lorentzians of width $\Delta$, one centered in
$\omega=\omega_0$ and one in $\omega= - \omega_0$:
\be
g(\omega) = \frac{\Delta}{2\pi}\left[ \frac{1}{\left(\omega - \omega_0 \right)^2 + \Delta^2}
+\frac{1}{\left( \omega + \omega_0 \right)^2 + \Delta^2} \right] \, ,
\label{lorendist}
\ee
which is normalized, $\int_{-\infty}^{+\infty} \dd \omega \, g(\omega) =1$. It is actually bimodal only if
$\omega_0 > \frac{\Delta}{\sqrt{3}}$. If this condition is not
satisfied, $g(\omega)$ is symmetric and unimodal, in which case the known results \cite{strogatz2000,reviewjstat}
do not show the richness of different behaviours that is found for bimodal distributions\footnote{We remind that usually
one considers frequency distributions centered in $\omega = 0$. This is not a loss of generality, since with a distribution
centered at any given value $\omega^*$, it is possible to perform a change of variables $\theta_i \to \theta_i +\omega^*t$,
going back to the former case.}.

In the $N \to \infty$ limit the dynamics can be described by the following Fokker-Planck equation for the
single particle distribution function $\rho (\theta,\omega,t)$:
\be
\frac{\partial}{\partial t}\rho(\theta,\omega,t) = -\frac{\partial}{\partial \theta}
\left[\left(\omega + F(\theta,t)\right)\rho(\theta,\omega,t)\right] +
D \frac{\partial^2}{\partial \theta^2}\rho(\theta,\omega,t) \, ,
\label{fpequat}
\ee
where $\rho(\theta,\omega,t)\dd \theta \dd \omega$ gives the fraction of oscillators with natural frequencies in the range
$(\omega,\omega +\dd \omega)$ that at time $t$ have phases in the range $(\theta,\theta + \dd \theta)$.
The distribution $\rho(\theta,\omega,t)$ is normalized for each $\omega$, i.e., $\int_0^{2\pi} \dd \theta \,
\rho(\theta,\omega,t) =1$, normalization which is conserved by the Fokker-Planck equation. We see that
actually Eq. (\ref{fpequat}) is a system of partial differential equations, one for each $\omega$, which are coupled by the
force term $F(\theta,t)$, given by  
\be
F(\theta,t) = K\int_{-\infty}^{+\infty} \dd \omega \, \int_0^{2\pi} \dd \theta' g(\omega) \sin (\theta' - \theta)
\rho(\theta',\omega,t) \, .
\label{forceeq}
\ee
The degree of synchronization of the system is best described by the complex order parameter $r(t)$, defined by
\be
r(t) = \int_{-\infty}^{+\infty} \dd \omega \, \int_0^{2\pi} \dd \theta g(\omega) e^{\ii \theta}
\rho(\theta,\omega,t) \, .
\label{deforder}
\ee
From Eqs. (\ref{forceeq}) and (\ref{deforder}) we see that $F(\theta,t) = K{\rm Im}\left[ r(t) e^{-\ii \theta}\right]$.
The order parameter satisfies $|r(t)| \le 1$. In a incoherent state we have $|r|=0$, while a fully synchronized state
has $|r|=1$. Using the order parameter in the Fokker-Planck equation (\ref{fpequat}), we obtain the expression which is
useful for the following analysis:
\be
\!\!\!\!\!\!\!\!\!\!\!\!\!\!\!\!\!\!\!\!\!\!\!\!\!\!\!\!\!\!\!
\frac{\partial}{\partial t}\rho(\theta,\omega,t) = -\frac{\partial}{\partial \theta}
\left[\left(\omega + \frac{K}{2\ii}(r(t)e^{-\ii \theta} - r^*(t)e^{\ii \theta})\right)\rho(\theta,\omega,t)\right] +
D \frac{\partial^2}{\partial \theta^2}\rho(\theta,\omega,t) \, ,
\label{fpstart}
\ee
where, as ususal, the star denotes complex conjugation. A Fourier expansion of the distribution function gives:
\be
\rho(\theta,\omega,t) = \frac{1}{2\pi} \sum_{n=-\infty}^{+\infty} f_n(\omega,t) e^{\ii n\theta} \, .
\label{fourexp}
\ee
The normalization and the reality of $\rho(\theta,\omega,t)$ imply that $f_0(\omega,t) \equiv 1$ and
$f_{-n}(\omega,t) = f_n^*(\omega,t)$. Susbstituting the Fourier expansion in the Fokker-Planck equation (\ref{fpstart}) we obtain
the following system of differential equations:
\bea
\dot{f}_n(\omega,t) &\equiv& \frac{\partial f_n(\omega,t)}{\partial t} = -\ii n \omega f_n(\omega,t) \nonumber \\
&&-\frac{K}{2}n\left[ r(t) f_{n+1}(\omega,t) - r^*(t) f_{n-1}(\omega,t)\right] -Dn^2f_n(\omega,t) \, .
\label{systemeq}
\eea
We see that the equation for $n=0$ gives $\dot{f}_0 = 0$, coherently with the fact that $f_0 \equiv 1$, and that the two open subsystems
for $n>0$ and $n<0$ are decoupled; however, the subsystem for $n<0$ is simply the complex conjugate of that for $n>0$, since $f_{-n}=f_n^*$,
and it is not necessary to consider it.

When $D=0$, the Ott-Antonsen ansatz consists in assuming that $f_n(\omega,t) = f_1^n(\omega,t)$ for each $n$ \cite{ott-anton,martens2009}. One of the
physical justification for the ansatz is that the known forms of the stationary states of the Kuramoto model, both for the incoherent and for the
synchronized case, satisfy the ansatz. It is easy to see that, plugging the ansatz
in each one of the Eqs. (\ref{systemeq}), they all become equal to the equation for $f_1(\omega,t)$. The presence of the noise term, i.e. the last
term of each equation of the system, does not allow to make the ansatz;
this is also consistent with the fact that the stationary solution of Eq. (\ref{fpstart}) does not satisfy it \cite{saka,reviewjstat}. However,
as we will show, in spite of this, the frequency distribution (\ref{lorendist}) makes it possible to perform an analysis of the dynamics of the order
parameter (\ref{deforder}), and to obtain a full phase diagram. As it will be clear in the following, this analysis is possible when the frequency
distribution can be analytically continued in the complex $\omega$ plane and it vanishes in the whole lower half plane when $|\omega| \to \infty$; this
is not verified, e.g., for a Gaussian frequency distribution. It will also be clear that the analysis is practically feasible when the number of poles
of the distribution is small.

After having obtained, for $D=0$, a closed equation for $f_1(\omega,t)$, in the successive analysis it is assumed that
$f_1(\omega,t)$ has no singularities in the lower half plane, and that
$f_1(\omega,t) \to 0$ for ${\rm Im}(\omega) \to -\infty$, the latter being based on the fact that it holds for any $t>0$ if it holds for $t=0$ \cite{martens2009}.
In our case we have to keep all the Fourier terms $f_n$, and we make the analogous assumptions that $f_n(\omega,t)$ do not have singularities
in the lower half plane and that $f_n(\omega,t) \to 0$ for
${\rm Im}(\omega) \to -\infty$. These assumptions can be justified as follows. We first note that the system of equations (\ref{systemeq}) can be continued
to the complex $\omega$ plane\footnote{If we consider the system for complex $\omega$, then it is no more true that the equations for negative $n$ are
the complex conjugates of those for positive $n$, and in fact for complex $\omega$ we cannot consider anymore $\rho(\theta,\omega,t)$ to be real,
and that $f_{-n}=f_n^*$. However,
this is not relevant, since, as we have seen, the subsysytems for negative and positive $n$ are decoupled, and for our analysis of the dynamics of the order
parameter (\ref{deforder}) we need only the equations with positive $n$.}. Second, denoting a complex $\omega$ with $\omega_{{\rm R}} + \ii \omega_{{\rm I}}$,
for large negative $\omega_{{\rm I}}$ the equation for $f_n$ can be approximated with $\dot{f}_n(\omega,t) = - |\omega_{{\rm I}}|f_n(\omega,t)$, showing that
for $t>0$ we have $f_n(\omega,t) \to 0$ when $\omega_{{\rm I}} \to -\infty$, if this holds for $t=0$.

Since we have to study the whole system of equations for positive $n$, we introduce the generalized complex order parameters:
\be
r_n(t) = \int_{-\infty}^{+\infty} \dd \omega \, \int_0^{2\pi} \dd \theta g(\omega) e^{\ii n \theta}
\rho(\theta,\omega,t) \, .
\label{defgenorder}
\ee
We see that the usual order parameter $r(t)$ is given by $r_1(t)$. Substituing the Fourier expansion for $\rho(\theta,\omega,t)$ we find
\be
r_n(t)= \int_{-\infty}^{+\infty} \dd \omega \, g(\omega) f_{-n}(\omega,t) \, ,
\label{orderfour_1}
\ee
or analogously
\be
r_n^*(t)= \int_{-\infty}^{+\infty} \dd \omega \, g(\omega) f_n(\omega,t) \, ,
\label{orderfour_2}
\ee
At this point one can make use of the above properties of the frequency distribution $g(\omega)$. Suppose that this function has $q$
poles in the lower half plane, denoted by $\omega_1,\dots,\omega_q$, and that it vanishes, in this half plane, when $|\omega|\to \infty$. Then,
from Eq. (\ref{orderfour_2}) we obtain
\be
r_n^*(t) = - 2\pi \ii \sum_{s=1}^q {\rm Res}\left. \left[g(\omega)f_n(\omega,t)\right] \right|_{\omega=\omega_s} \, ,
\label{residues}
\ee
i.e., $r_n^*(t)$ is proportional to the sum of the residues of the function $g(\omega)f_n(\omega,t)$ computed at the poles of $g(\omega)$.
The poles in the lower half plane of the function $g(\omega)$ in Eq. (\ref{lorendist}) are in $\omega = \omega_0 - \ii \Delta$ and in
$\omega = -\omega_0 -\ii \Delta$, and are simple. Then, from eq. (\ref{residues}) we get
\be
r_n^*(t) = \frac{1}{2} f_n(\omega_0 -\ii \Delta) + \frac{1}{2} f_n(-\omega_0 - \ii \Delta) \, .
\label{exprrstar}
\ee
At this point we can define
\bea
\label{exprr1star}
r_n^{(1)*}(t) &=& f_n(\omega_0 - \ii \Delta) \\
\label{exprr2star}
r_n^{(2)*}(t) &=& f_n(-\omega_0 - \ii \Delta) \, ,
\eea
so that
\be
r_n(t) = \frac{1}{2} \left[ r_n^{(1)}(t) + r_n^{(2)}(t)\right]
\label{sumr}
\ee
From the equations (\ref{systemeq}) we thus have:
\bea
\label{eqrn1star}
\!\!\!\!\!\!\!\!\!\!\!\!\!\!\!\!\!\!\!\!\!\!\!\!\!\!\!\!\!\!\!
\dot{r}_n^{(1)*} &=& -n\left( \Delta +\ii \omega_0 +nD\right)r_n^{(1)*} -\frac{K}{4}n\left[ \left( r_1^{(1)}
+r_1^{(2)} \right) r_{n+1}^{(1)*} - \left( r_1^{(1)*} + r_1^{(2)*} \right) r_{n-1}^{(1)*} \right] \\
\!\!\!\!\!\!\!\!\!\!\!\!\!\!\!\!\!\!\!\!\!\!\!\!\!\!\!\!\!\!\!
\label{eqrn2star}
\dot{r}_n^{(2)*} &=& -n\left( \Delta -\ii \omega_0 +nD\right)r_n^{(2)*} -\frac{K}{4}n\left[ \left( r_1^{(1)}
+r_1^{(2)} \right) r_{n+1}^{(2)*} - \left( r_1^{(1)*} + r_1^{(2)*} \right) r_{n-1}^{(2)*} \right] \, ,
\eea
i.e.,
\bea
\label{eqrn1}
\!\!\!\!\!\!\!\!\!\!\!\!\!\!\!\!\!\!\!\!\!\!\!\!\!\!\!\!\!\!\!
\dot{r}_n^{(1)} &=& -n\left( \Delta -\ii \omega_0 +nD\right)r_n^{(1)} -\frac{K}{4}n\left[ \left( r_1^{(1)*}
+r_1^{(2)*} \right) r_{n+1}^{(1)} - \left( r_1^{(1)} + r_1^{(2)} \right) r_{n-1}^{(1)} \right] \\
\!\!\!\!\!\!\!\!\!\!\!\!\!\!\!\!\!\!\!\!\!\!\!\!\!\!\!\!\!\!\!
\label{eqrn2}
\dot{r}_n^{(2)} &=& -n\left( \Delta +\ii \omega_0 +nD\right)r_n^{(2)} -\frac{K}{4}n\left[ \left( r_1^{(1)*}
+r_1^{(2)*} \right) r_{n+1}^{(2)} - \left( r_1^{(1)} + r_1^{(2)} \right) r_{n-1}^{(2)} \right] \, .
\eea
Eqs. (\ref{eqrn1}) and (\ref{eqrn2}), for $n=1,2,\dots$, or better their dimensionless version to be introduced shortly,
are the basic equations of our study. In the equation for $n=1$ there appear
$r_0^{(1)}$ and $r_0^{(2)}$, that are understood to be identically equal to $1$.  We see that from the original Fokker-Planck equation
(\ref{fpstart}), where we had an infinite dimensional dynamical system labelled by two continuous variables $\theta$ and $\omega$,
we have obtained a dynamical system which is still infinite dimensional, but labelled by only the discrete variable $n$. This has been
made possible by the above mentioned properties of the frequency distribution $g(\omega)$, and it is due to the fact that we restrict our interest
on the dynamics of the order parameters. We emphasize here two things: the order parameters, and in particular $r_1$, are the most relevant
quantities in our system of interacting oscillators, on which the characterization of the properties of
the asymptotic states are based. Second, we repeat that without the above properties of $g(\omega)$ one could not go beyond
Eqs. (\ref{systemeq}), so that a system of discrete differential equations for the order parameters could not be written\footnote{For general
$g(\omega)$ it is possible to write a system of linear differential equations, system labelled by the continuous variable $\omega$, that can be solved
with the Laplace transform, but this analysis is restricted to the study of the dynamics of a vanishingly small order parameter $r_1(t)$, i.e., to the
stability analysis of the incoherent state $r_1=0$ \cite{strogatz2000}.}.

\subsection{Reduced variables}
\label{redvar}

The system can conveniently be studied using dimensionless parameters; this also allows an easier comparison with the results
of the noiseless ($D=0$) system \cite{martens2009}. This can be achieved by defining the quantities $\widehat{\Delta} = \frac{4}{K}\Delta$,
$\widehat{\omega}_0 = \frac{4}{K}\omega_0$, $\widehat{D} = \frac{4}{K}D$, and $\widehat{t} = \frac{4}{K}t$. Eqs. (\ref{eqrn1}) and (\ref{eqrn2}) now read
\bea
\label{eqrn1h}
\!\!\!\!\!\!\!\!\!\!\!\!\!\!\!\!\!\!\!\!\!\!\!\!\!\!\!\!\!\!\!
\dot{r}_n^{(1)} &=& -n\left( \widehat{\Delta} -\ii \widehat{\omega}_0 +n\widehat{D}\right)r_n^{(1)} -n\left[ \left( r_1^{(1)*}
+r_1^{(2)*} \right) r_{n+1}^{(1)} - \left( r_1^{(1)} + r_1^{(2)} \right) r_{n-1}^{(1)} \right] \\
\!\!\!\!\!\!\!\!\!\!\!\!\!\!\!\!\!\!\!\!\!\!\!\!\!\!\!\!\!\!\!
\label{eqrn2h}
\dot{r}_n^{(2)} &=& -n\left( \widehat{\Delta} +\ii \widehat{\omega}_0 +n\widehat{D}\right)r_n^{(2)} -n\left[ \left( r_1^{(1)*}
+r_1^{(2)*} \right) r_{n+1}^{(2)} - \left( r_1^{(1)} + r_1^{(2)} \right) r_{n-1}^{(2)} \right] \, ,
\eea
where now the dot denotes differentiation with respect to $\widehat{t}$.
In the following, to ease the notation we will drop the hat over the dimensionless parameters. Only in Section \ref{simdyn}, presenting the results of
numerical simulations of the full system of equations (\ref{langeq}), we will reintroduce the use of the hat, since we will have to refer also to the
original system parameters.

We are interested in the asymptotic solutions of the system of equations. As we will see, these are given either by stable stationary states, or by
standing wave states. We note that the equations are invariant with respect to the transformation $r_n^{(1)} \to r_n^{(1)}e^{-\ii n \psi}$
and $r_n^{(2)} \to r_n^{(2)}e^{-\ii n \psi}$ for arbitrary $\psi$; this is a consequence of the global rotational invariance of the system.

\section{The stationary states}
\label{statstates}

\subsection{The incoherent state and its stability}
\label{incohstate}

For any value of the parameters, the incoherent state $r_n^{(1)} = r_n^{(2)} = 0$ for any $n$ is a solution of the system of equations. We begin our
analysis by studying the stability of the incoherent solution. By linearizing Eqs. (\ref{eqrn1h}) and (\ref{eqrn2h}) we obtain:
\bea
\label{eqrn1hsa}
\!\!\!\!\!\!\!\!\!\!\!\!\!\!\!\!\!\!\!
\dot{r}_1^{(1)} &=& -\left( \Delta -\ii \omega_0 +D\right)r_1^{(1)} + ( r_1^{(1)} + r_1^{(2)} ) \\
\!\!\!\!\!\!\!\!\!\!\!\!\!\!\!\!\!\!\!
\label{eqrn2hsa}
\dot{r}_1^{(2)} &=& -\left( \Delta +\ii \omega_0 +D\right)r_1^{(2)} + ( r_1^{(1)} + r_1^{(2)} )  \\
\label{eqrn1hsb}
\!\!\!\!\!\!\!\!\!\!\!\!\!\!\!\!\!\!\!
\dot{r}_n^{(1)} &=& -n\left( \Delta -\ii \omega_0 +nD\right)r_n^{(1)} \,\,\,\,\,\,\,\,\,\,\,\,\,\,\,\, n>1 \\
\!\!\!\!\!\!\!\!\!\!\!\!\!\!\!\!\!\!\!
\label{eqrn2hsb}
\dot{r}_n^{(2)} &=& -n\left( \Delta +\ii \omega_0 +nD\right)r_n^{(2)} \,\,\,\,\,\,\,\,\,\,\,\,\,\,\,\, n>1 \, ,
\eea
The incoherent sate is stable when all the eigenvalues of this linear system have a negative real part. The equations
for $n>1$ are all decoupled, and they directly give the eigenvalues, all of which with negative real part equal to $-n\Delta -n^2 D$.
The two equations for $n=1$, on the other hand, are coupled, and a simple calculation shows that the corresponding eigenvalues are
given by:
\be
\label{eigen}
\lambda = -\left( \Delta  + D -1 \right) \pm \sqrt{1-\omega_0^2} 
\ee
The phase diagram will be considered, for any given fixed value of $D$, in the first quadrant of the $(\omega_0,\Delta)$ plane (i.e., the quadrant
where both parameters are positive, the meaningful case). Then, let us see
which are, in this quadrant, the boundaries defining the regions where the incoherent state is stable. We have to distinguish the cases
$\omega_0 \le 1$ and $\omega_0 > 1$. In the former case we obtain that stability requires $\Delta > 1 - D +\sqrt{1-\omega_0^2}$,
since both eigenvalues must be negative,
while in the latter case it must be $\Delta > 1 - D$ (then if $D\ge1$ any positive $\Delta$ satisifies stability for $\omega_0 >1$).
In the former case the inequality defines, in the first quadrant of the $(\omega_0,\Delta)$ plane, the part of the strip $0\le \omega_0 \le 1$
which is above the circle with center in $(\omega_0,\Delta) = (0,1-D)$ and radius equal to $1$. When $D\ge 2$ this region coincides with the entire strip;
thus, for $D\ge 2$ the incoherent state is always stable. We will see in the following that, actually, for $D\ge 2$ the phase
diagram is trivial, since the only stationary state is the incoherent state; thus, the interesting cases occur for $D<2$.

\subsection{Partially sinchronized stationary states}
\label{partsynch}

Eqs. (\ref{eqrn1h}) and (\ref{eqrn2h}) admit other stationary solutions. In the noiseless $D=0$ case the Ott-Antonsen ansatz allows to have
a close system involving just $r_1^{(1)}$ and $r_1^{(2)}$, and it is possible to perform an analytical evaluation of the stationary solutions.
In our case we have an open system, and we have to resort to a numerical evaluation. This can be done by truncating the system to a given value
of $n$, sufficiently large to represent with very good approximation the full infinite system. That this truncation is a feasible and meaningful approximation
can be understood from the fact that it is expected that a stationary state will give rise to a distribution function $\rho(\theta,\omega)$ such that
the modulus of the order parameters $r_n$ decreases rapidly with $n$, since to have a finite value of $r_n$ for large $n$ requires a wildly fluctuating
distribution function. Therefore, our numerical study of the system of equations has been performed by putting equal to $0$ all the variables
$r_n^{(1)}$ and $r_n^{(2)}$ for $n>M$, thus obtaining a closed system of $2M$ equations. We have chosen $M=50$, and we have verified that in all cases
our results for the main order parameter $r_1$ do not change by taking a larger value of $M$.

\begin{figure}[h]
\begin{center}
\includegraphics[scale=0.5,trim= 0cm 3.0cm 0cm 0cm,clip]{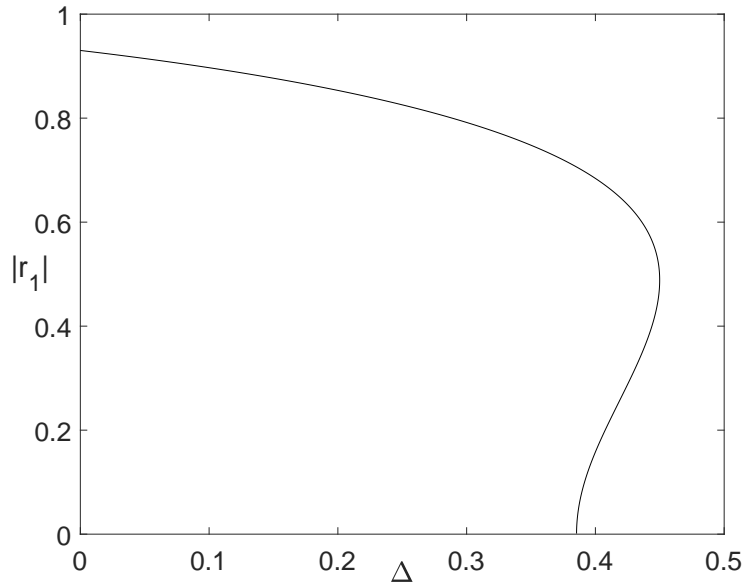}
\caption{The order parameter $|r_1|$ corresponding to the stationary state of the system of equations (\ref{eqrn1h}) and (\ref{eqrn2h}) as a function of
$\Delta$ for $D=0.5$, with $\omega_0$ constrained to be equal to $2.58\Delta$.}
\label{fig_r_of_s}
\end{center}
\end{figure}

In Fig. \ref{fig_r_of_s} we plot an example of the value of the modulus $|r_1|$ of the order parameter $r_1 = \frac{1}{2}(r_1^{(1)}+r_1^{(2)})$
corresponding to the stationary state of the system of
equations (\ref{eqrn1h}) and (\ref{eqrn2h}).
This quantity is plotted for the particular value $D=0.5$ and as a function of $\Delta$, with $\omega_0$ constrained to be equal to $2.58 \Delta$;
practically, $|r_1|$ is given as a function of a quantity proportional to the distance from the origin in the first quadrant of the $(\omega_0,\Delta)$,
distance computed along the line $\omega_0 = 2.58 \Delta$.
There is no reason for the choice of a particular value of the latter proportionality constant, i.e. we could have chosen another proportionality constant
$\gamma$ such that $\omega_0 = \gamma \Delta$. The only important thing is that, in order to have the
structure shown in the plot, with $|r_1|$ initially increasing from $0$ for increasing $\Delta$ (i.e., with the curve starting towards the right), $\gamma$
must be larger than a $D$ dependent value, that we now consider. It is known \cite{strogatz2000} that for $D=0$ the syncronization transition is supercritical
for a symmetric unimodal frequency distribution $g(\omega)$, while it is subcritical for a symmetric bimodal distribution, in which the second derivative
of $g(\omega)$ at $\omega =0$ is positive. In the latter case we would have a plot qualitatively similar to the one in Fig. \ref{fig_r_of_s}, while in the former
case the curve would start towards the left. For our form of $g(\omega)$ the bimodal case corresponds to $\omega_0> \frac{\Delta}{\sqrt{3}}$. This picture holds
also in the noisy case $D>0$, but the discriminating value of $\omega_0$ depends on $D$. This value can be found using a power series expansion that gives
the stationary value of $r_1$ as a function of the parameters of the distribution $g(\omega)$; one obtains an expression valid for
$r_1 \to 0$, sufficient to see if the synchronization transition is supercritical or subcritical \cite{kurabrief}. The general expressions give
the critical value $K_c$ of the coupling constant $K$ where the synchronization transition occurs and the value of $r_1$ for $K$ in the neighborhood of
$K_c$ \cite{kurabrief}; using our dimensionless parameters one finds, for the particular $g(\omega)$ given by Eq. (\ref{lorendist}), that the synchronization
transition occurs for
\begin{equation}
\label{synchtrans}
\omega_0^2 = 2( \Delta + D) - (\Delta + D)^2 \, ,
\end{equation}
and that the transition is subcritical if
\begin{equation}
\label{synchsub}
\omega_0^2 > (\Delta + 2D)(\Delta + D)^2/(3\Delta + 4D) \, .
\end{equation}
In the $(\omega_0,\Delta)$ plane, expression (\ref{synchtrans}) defines the circumference with center in $(\omega_0,\Delta)=(0,1-D)$ and radius equal to $1$.
At the transition points defined by Eq. (\ref{synchtrans}) a stationary solution with positive $|r_1|$ bifurcates continuously from the incoherent
solution $r_1=0$.
When the transition is subcritical and thus the plot of $|r_1|$ {\it vs} $\Delta$ is as in
Fig. \ref{fig_r_of_s}, we see that, for the range of $\Delta$ between that of the transition and the one where the curve reaches its rightmost point,
there are two stationary states with positive $|r_1|$ (beyond the one with $r_1=0$ which is always present). We expect that the one corresponding to the larger
value of $r_1$ is stable, while the other is unstable, and this has been confirmed in all cases by the numerical results. 
It is easy to see that the $\omega_0$ value of the transition, as given in eq. (\ref{synchtrans}), is always $\le 1$, and that solved
for $\Delta$ it gives $\Delta = 1 - D \pm \sqrt{1-\omega_0^2}$; the solution with the positive sign before the square root coincides with the threshold for the
stability of the incoherent state.

\begin{figure}[h]
\begin{center}
\includegraphics[scale=0.5,trim= 0cm 3.0cm 0cm 0cm,clip]{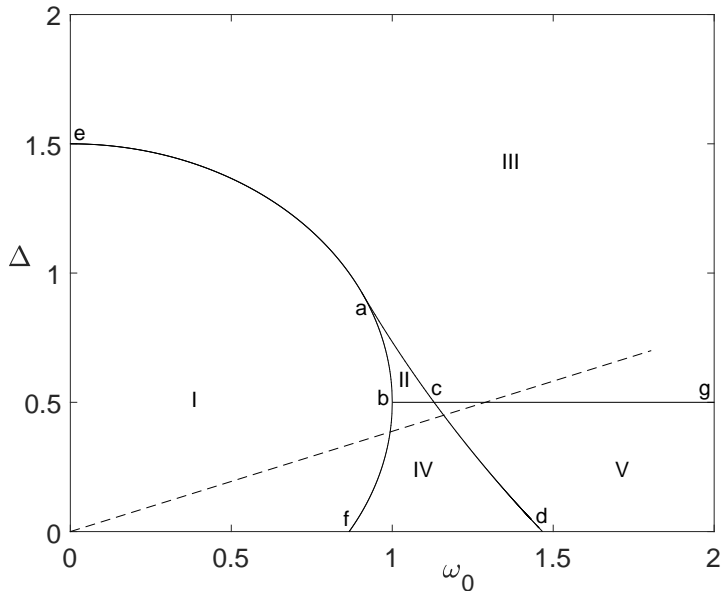}
\caption{The partial structure of the phase diagram for $D=0.5$. See the text for the description of the full lines, the meaning of the lowercase letters
and roman numbers, and an explanation of the stability properties of
the stationary states in the different regions of the diagram. The dashed line is the one along which the plot of $|r_1|$ {\it vs} $\Delta$ in
Fig. \ref{fig_r_of_s} has been computed.}
\label{fig_part_struct}
\end{center}
\end{figure}

The above analysis allows us to build a partial (but almost complete) structure of the phase diagram, that we plot in Fig. \ref{fig_part_struct} for $D=0.5$,
the same value used in
Fig. \ref{fig_r_of_s}. The structure will be completed in the next section after considering the standing wave states. We prefer to present at this point
the picture of the phase diagram, to show its richness, since the further feature that will be introduced after the analysis in section \ref{wavestates}
involves only a relatively small region of the diagram.
  
In the plot lowercase letters are used to denote the intersections of different full lines or of a full line with one coordinate axis or a boundary of the plot;
roman numbers are used to denote the
different regions delimited by these lines. The curved line {\it `eabf'}
is the part of the circumference with center in $(\omega_0,\Delta)=(0,1-D)=(0,0.5)$, and radius equal to $1$, that lies in the first quadrant;
the horizontal line {\it `bcg'}, actually extending indefinitely for $\omega_0>1$, is at $\Delta=1-D=0.5$, while the other almost (but not quite) straight line
{\it `acd'} is the locus of $(\omega_0,\Delta)$
points where the plots like the one on Fig. \ref{fig_r_of_s} reach their righmost point, each plot characterized by a different $\gamma$ value. The point
{\it `a'}, intersection of this line with the circumference, is the one where Eq. (\ref{synchsub}) is satisfied as an equality, in addition to Eq. (\ref{synchtrans}).
According to
our analysis, the incoherent state $r_1=0$ is stable in the part of the diagram which is, at the same time, outside the circumference and above the
horizontal line, i.e., in regions II and III; on the other hand, the partially synchronized state exists and is stable inside the circumference and in the 
part of the diagram outside the circumference but
to the left of the line {\it `acd'}, i.e., in regions I, II and IV. Then in the small region II, above the horizontal line and delimited by the circumference
and the line {\it `acd'} (i.e., the region {\it `abc'}), both the incoherent state and the partially synchronized state are stable; thus the stationary state
reached by the system depends on the initial conditions. We have here the first example of bi-stability.

We note the following. As remarked above, at the points defined by Eq. (\ref{synchtrans}) a stationary solution with positive $|r_1|$ bifurcates from the
incoherent solution. If this transition is supercritical, the bifurcating solution is the only one with positive $|r_1|$, and it is stable, while the incoherent
solution is stable up to the transition point; this occurs in the section {\it `ea'} of the circumference, with the incoherent solution stable outside
the circumference and up to it, and the solution with positive $|r_1|$ existing and stable inside the circumference. On the other hand, when the transition is
subcritical, as in the section {\it `abf'} of the circumference, at the transition the positive $|r_1|$ solution bifurcating from the incoherent one is unstable,
and there is another partially synchronized state with larger $|r_1|$. As far as the stability of the incoherent state is concerned, we have seen above, in Section
\ref{incohstate}, that it is stable outside and up to the circumference for $\Delta > 1-D$, while it is already unstable, before reaching the circumference,
for $\Delta < 1-D$ (see also below, Section \ref{classbif}).

Finally, the dashed line is the one along which the plot of $|r_1|$ {\it vs} $\Delta$ in
Fig. \ref{fig_r_of_s} has been computed.

Region V is the one where standing wave states occur. We will study these states in section \ref{wavestates}, and we will complete the phase diagram
for $D=0.5$. Afterwards, we will consider
other ranges for the value of $D$, where the structure of the diagram changes. But for the moment, we treat some features related to the stability of the
partially synchonized stationary states.

\subsection{Stability properties}
\label{stabprop}

In all cases we have found that the partially synchronized stationary states are represented by stationary solutions of Eqs. ({\ref{eqrn1h})
and (\ref{eqrn2h}) where $|r_n^{(1)}| = |r_n^{(2)}|$, in particular $|r_1^{(1)}|=|r_1^{(2)}|$. This had to be expected on physical grounds, from the
invariance of the frequency distribution $g(\omega)$ for $\omega \to -\omega$. The numerical solutions of Eqs. ({\ref{eqrn1h}) and (\ref{eqrn2h})
have shown that these aymptotic solutions are reached even when the initial conditions do not satisfy the above equalities. Although without
any rigour, this should practically prove that there are not stationary states that do not satisfy the equalitites.

In this section we study numerically the linear stability of the partially synchronized states, and for this purpose we use a modified version of the
systems of equations. Precisely, we go from the variables $r_n^{(1)}$ and $r_n^{(2)}$ to the variables $r_n$ already defined in Eq. (\ref{sumr})
and new variables $b_n$:
\bea
\label{sumr1}
r_n &=& \frac{1}{2} \left[ r_n^{(1)} + r_n^{(2)}\right] \\
\label{sumr2}
b_n &=& \frac{1}{2\ii} \left[ r_n^{(1)} - r_n^{(2)}\right] \, .
\eea
With these variables Eqs. (\ref{eqrn1h}) and (\ref{eqrn2h}) are replaced by:
\bea
\label{eqrnah}
\!\!\!\!\!\!\!\!\!\!\!\!\!\!
\dot{r}_n &=& -n\left( \Delta +nD\right)r_n -n\omega_0 b_n -2n \left( r_1^* r_{n+1} - r_1 r_{n-1} \right) \\
\!\!\!\!\!\!\!\!\!\!\!\!\!\!
\label{eqrnbh}
\dot{b}_n &=& -n\left( \Delta +nD\right)b_n +n\omega_0 r_n -2n \left( r_1^* b_{n+1} - r_1 b_{n-1} \right) \, .
\eea
In the equations for $n=1$ it is understood that $r_0 \equiv 1$ and $b_0 \equiv 0$.
Like Eqs. (\ref{eqrn1h}) and (\ref{eqrn2h}), these equations are invariant with respect to the transformation $r_n \to r_ne^{-\ii n \psi}$
and $b_n \to b_ne^{-\ii n \psi}$ for arbitrary $\psi$, due to the global rotational invariance of the system.
The coefficients appearing in these equations are all real, and then it is possible to restrict the study to real solutions. From eqs. (\ref{sumr1})
and (\ref{sumr2}) we see that real solutions $r_n$ and $b_n$ require not only that $|r_n^{(1)}| = |r_n^{(2)}|$, but in addition also that
$r_n^{(1)*} = r_n^{(2)}$. However, the transformations $r_n \to r_ne^{-\ii n \psi}$ and $b_n \to b_ne^{-\ii n \psi}$, that brings solutions
into solutions, while spoiling the latter equality, conserves the former, that, as mentioned above, has always been found to be satisfied
for the partially synchronized stationary states. This suggests to study the linear stability of the partially synchronized states
in the following way (although the numerical results clearly show that these states are stable, it is useful to have an independent confirmation
based on a linear stability analysis). We consider a real solution of Eqs. (\ref{eqrnah}) and (\ref{eqrnbh}) corresponding to such a state, and we
linearize the equations with respect to this solution. Then we can study separately the real and the imaginary parts of the linearized equations.

Denoting $r_n(t) = r_n^0 + \delta r_n(t)$ and $b_n(t) = b_n^0 +\delta b_n(t)$, where $r_n^0$ and $b_n^0$ are the real stationary solutions,
the linearized equations are:
\bea
\label{eqrnahlin}
\!\!\!\!\!\!\!\!\!\!\!\!\!\!\!\!\!\!\!\!\!\!\!\!\!\!\!\!\!\!\!\!\!\!
\dot{\delta r}_n &=& -n\left( \Delta +nD\right)\delta r_n -n\omega_0 \delta b_n -2n \left( \delta r_1^* r_{n+1}^0
+r_1^0 \delta r_{n+1} - \delta r_1 r_{n-1}^0 -r_1^0 \delta r_{n-1} \right) \\
\!\!\!\!\!\!\!\!\!\!\!\!\!\!\!\!\!\!\!\!\!\!\!\!\!\!\!\!\!\!\!\!\!\!
\label{eqrnbhlin}
\dot{\delta b}_n &=& -n\left( \Delta +nD\right)\delta b_n +n\omega_0 \delta r_n -2n \left( \delta r_1^* b_{n+1}^0
+r_1^0 \delta b_{n+1} - \delta r_1 b_{n-1}^0 -r_1^0 \delta b_{n-1} \right) \, .
\eea
Writing explicitly the real and the imaginary parts of the perturbations, $\delta r_n = \delta r_n^{\rm R} + \ii \delta r_n^{\rm I}$ and
$\delta b_n = \delta b_n^{\rm R} + \ii \delta b_n^{\rm I}$, we obtain the two sets of equations:
\bea
\label{eqrnahlinr}
\!\!\!\!\!\!\!\!\!\!\!\!\!\!\!\!\!\!\!\!\!\!\!\!\!\!\!\!\!\!\!\!\!\!
\dot{\delta r}_n^{\rm R} &=& -n\left( \Delta +nD\right)\delta r_n^{\rm R} -n\omega_0 \delta b_n^{\rm R} -2n \left( \delta r_1^{\rm R} r_{n+1}^0
+r_1^0 \delta r_{n+1}^{\rm R} - \delta r_1^{\rm R} r_{n-1}^0 -r_1^0 \delta r_{n-1}^{\rm R} \right) \\
\!\!\!\!\!\!\!\!\!\!\!\!\!\!\!\!\!\!\!\!\!\!\!\!\!\!\!\!\!\!\!\!\!\!
\label{eqrnbhlinr}
\dot{\delta b}_n^{\rm R} &=& -n\left( \Delta +nD\right)\delta b_n^{\rm R} +n\omega_0 \delta r_n^{\rm R} -2n \left( \delta r_1^{\rm R} b_{n+1}^0
+r_1^0 \delta b_{n+1}^{\rm R} - \delta r_1^{\rm R} b_{n-1}^0 -r_1^0 \delta b_{n-1}^{\rm R} \right)
\eea
for the real parts, and
\bea
\label{eqrnahlini}
\!\!\!\!\!\!\!\!\!\!\!\!\!\!\!\!\!\!\!\!\!\!\!\!\!\!\!\!\!\!\!\!\!\!
\dot{\delta r}_n^{\rm I} &=& -n\left( \Delta +nD\right)\delta r_n^{\rm I} -n\omega_0 \delta b_n^{\rm I} -2n \left( -\delta r_1^{\rm I} r_{n+1}^0
+r_1^0 \delta r_{n+1}^{\rm I} - \delta r_1^{\rm I} r_{n-1}^0 -r_1^0 \delta r_{n-1}^{\rm I} \right) \\
\!\!\!\!\!\!\!\!\!\!\!\!\!\!\!\!\!\!\!\!\!\!\!\!\!\!\!\!\!\!\!\!\!\!
\label{eqrnbhlini}
\dot{\delta b}_n^{\rm I} &=& -n\left( \Delta +nD\right)\delta b_n^{\rm I} +n\omega_0 \delta r_n^{\rm I} -2n \left( -\delta r_1^{\rm I} b_{n+1}^0
+r_1^0 \delta b_{n+1}^{\rm I} - \delta r_1^{\rm I} b_{n-1}^0 -r_1^0 \delta b_{n-1}^{\rm I} \right)
\eea
for the imaginary parts. Note that the two systems differ only for the sign of one term in each equation. 
In a transformation $r_n \to r_ne^{-\ii n\psi}$ and $b_n \to b_n e^{-\ii n \psi}$ with infinitesimal $\delta \psi$ we have, at
first order, $\delta r_n = -\ii n \delta \psi r_n$ and $\delta b_n = -\ii n \delta \psi b_n$. And actually we have found, for any real $r_n^0$ and $b_n^0$
corresponding to a stationary state, that the eigenvalues of the system (\ref{eqrnahlinr}) and (\ref{eqrnbhlinr}) have all negative real parts,
while the eigenvalues of the system (\ref{eqrnahlini}) and (\ref{eqrnbhlini}) have all negative real parts except one vanishingly eigenvalue, whose
eigenvector is exactly of the form $\delta r_n = -n \delta \psi r_n^0$ and $\delta b_n = -n \delta \psi b_n^0$. This confirms the expected
stability of the partially synchronized stationary state and the existence of a neutral perturbation corresponding to a global rotation of the system.

\section{The standing wave states and the full phase diagram}
\label{wavestates}

According to the above analysis, in the part of the diagram of Fig. \ref{fig_part_struct} which is below the horizontal line and to the right of the
line {\it `acd'}, i.e., in region V, a partially sinchronized state does not exist, and the incoherent state is not stable. Nevertheless the system reaches an
asymptotic state, that however is not a stationary state, but a periodic state, characterized by a limit cycle in the dynamical phase space of the system. At this point
we have to warn the reader that speaking of a limit cycle in our case is an abuse of mathematical terminology. For the noiselss $D=0$ case, studied
with the Ott-Antonsen ansatz, it was possible to use the concept of limit cycle because the presence of a closed system for $r_1^{(1)}(t)$ and $r_1^{(2)}(t)$,
and the assumption that $|r_1^{(1)}(t)|=|r_1^{(2)}(t)|$, reduced the study to a system of two equations \cite{martens2009}. In our case the dimensionality
of the dynamical phase space is, after the truncation, equal to $4M$ ($2M$ complex equations), in particular it is larger than $2$, and therefore it is not
possible to apply the Poincar\'e-Bendixson theorem. Then, our claim of a limit cycle is only a consequence of the numerical study of the
asymptotic state of the system of Eqs. (\ref{eqrn1h}) and (\ref{eqrn2h}), that appears to be periodic in region V of the phase diagram.

Given the above remark, we note that a periodic state, corresponding to the propagation of a standing wave in the system, appears to be physically very
reasonable. While the partial synchronization of a stationary state is realized when a macroscopic fraction of oscillators is locked to $\omega=0$, with the
remaining oscillators drifting, one can argue that a standing wave state is realized when there are two macroscopic fractions, each one locked
at a frequency close to that of one of the two peaks of $g(\omega)$, with the remaining oscillators drifting. By the symmetry of $g(\omega)$, it is
expected that the two locked frequencies are symmetric with respect to $\omega=0$, and that the two macroscopic frations are equally populated. Then, the two
groups of locked oscillators rotate in opposite directions.
In Fig. \ref{fig_stand_wave} we plot an example of the solution of the system of Eqs. (\ref{eqrn1h}) and (\ref{eqrn2h}) in a case where the asymptotic state
is  standing wave.
\begin{figure}[h]
\begin{center}
\includegraphics[scale=0.5,trim= 0cm 3.0cm 0cm 0cm,clip]{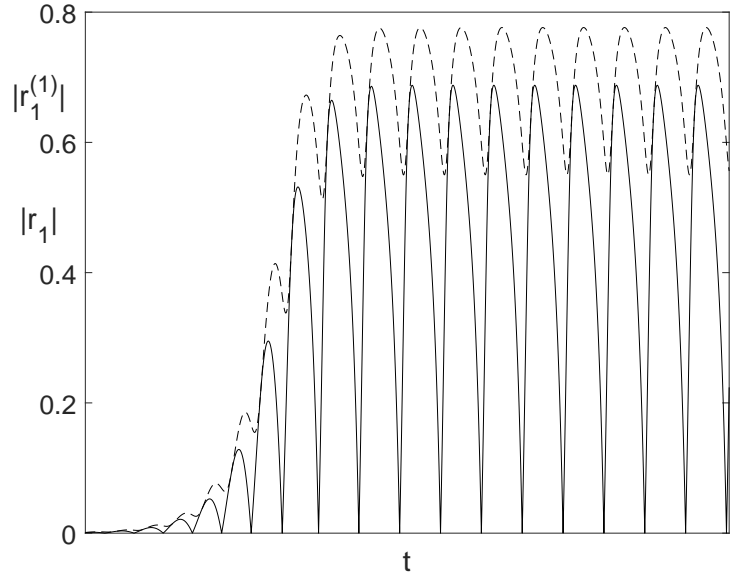}
\caption{The solution of the system of Eqs. (\ref{eqrn1h}) and (\ref{eqrn2h}) in a case in which the asymptotic state is a standing wave. The full line
shows the absolute value of the order parameter $r_1$, while the dashed line is the absolute value of $r_1^{(1)}$, that, once the asymptotic
standing wave state is reached, is equal to that of $r_1^{(2)}$.
Time is in arbitrary units.}
\label{fig_stand_wave}
\end{center}
\end{figure}
As for the stationary states, also for the standing wave states the equality $|r_1^{(1)}(t)|=|r_1^{(2)}(t)|$ is realized. The dashed line in the figure
shows $|r_1^{(1)}(t)|$. It should be now clear what happens: $r_1^{(1)}(t)$ and $r_1^{(2)}(t)$ rotate in opposite directions with nonconstant angular velocities
that are, at any $t$, equal in magnitude. When the angle between the two quantities in the complex plane is equal to $\pi$, i.e., when
$r_1^{(1)}(t)=-r_1^{(2)}(t)$, then $r_1(t)=r_1^{(1)}(t)+r_2^{(2)}(t)=0$, and this explains why in the standing wave state $|r_1(t)|$ reaches the value zero.

A study of the linear stability of the standing wave would require a Floquet analysis of the periodic solution of the system of equations.
We have not performed this analysis,
and we therefore limit ourselves to the numerical evidence of the stability of such states. 

Actually, it happens that standing wave states, that we remind are found as the only asymptotic states in region V, occur also in part of region IV, where
the incoherent state is unstable. This region is delimited by another line, the one denoted by {\it `bh'}, as shown in Fig. \ref{fig_comp_struct}.
This plot gives the complete phase diagram for $D=0.5$. Comparing with Fig. \ref{fig_part_struct} we note that part of the region denoted by IV in that figure,
now constitutes the new region denoted by VI. In the new region VI both the partially synchronized state and the standing wave state are stable,
the one reached as the asymptotic state depending on the initial conditions. We have here the other example of bi-stability.
\begin{figure}[h]
\begin{center}
\includegraphics[scale=0.5,trim= 0cm 3.0cm 0cm 0cm,clip]{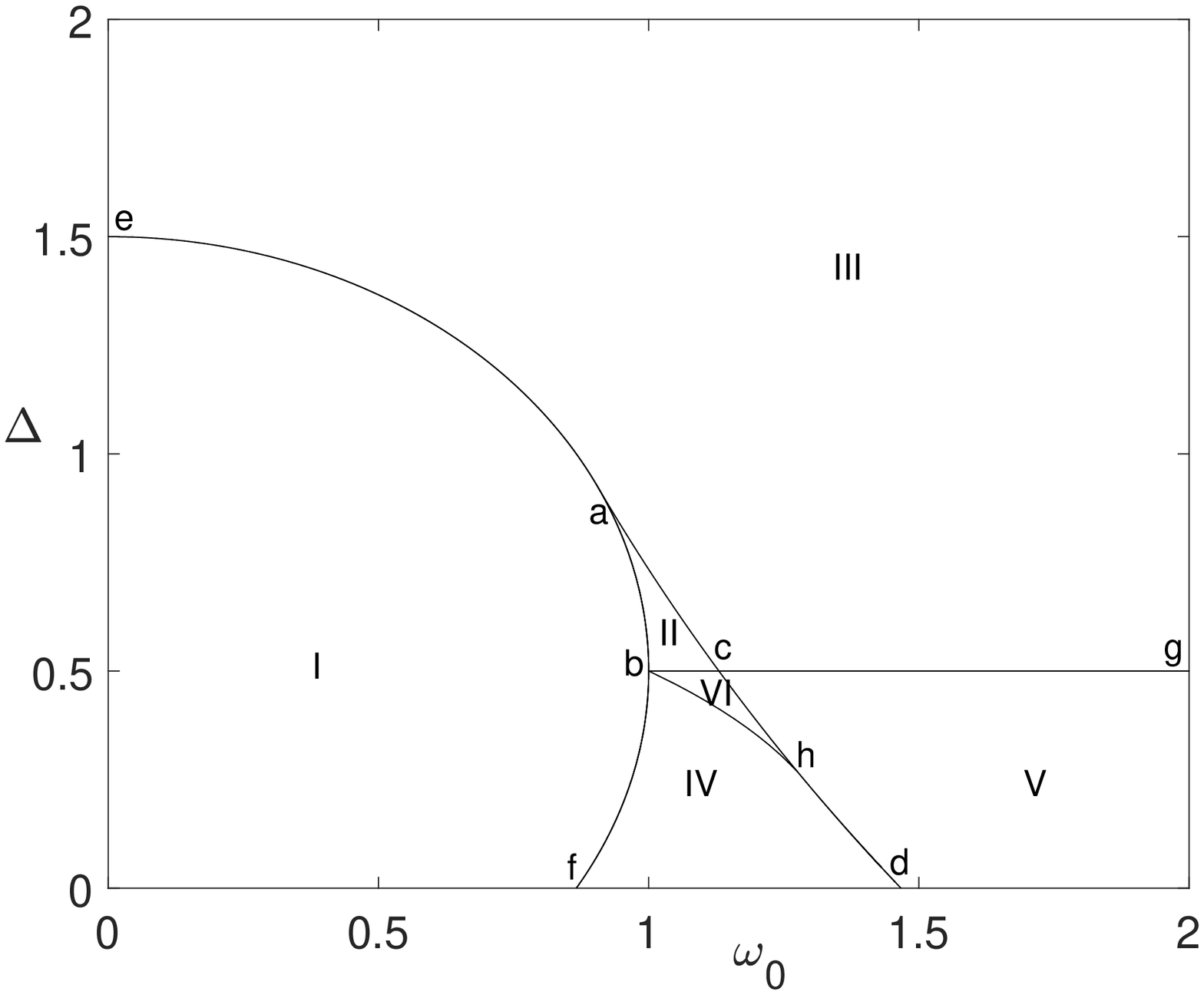}
\includegraphics[scale=0.5,trim= 0cm 3.0cm 0cm 0cm,clip]{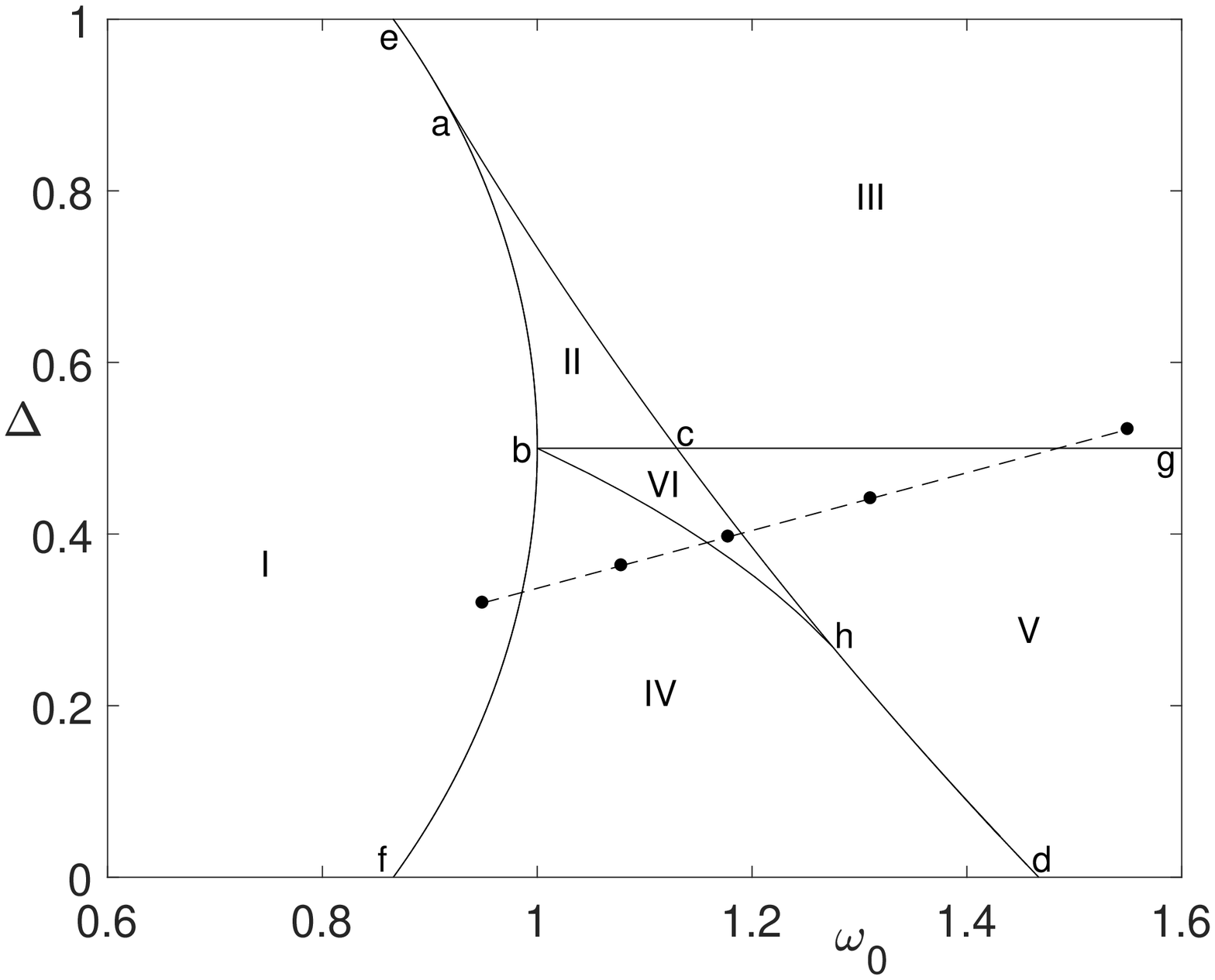}
\caption{The full phase diagram for $D=0.5$. The lower panel is a zoom of the more interesting region. As explained in the text, regions II and VI are characterized
by the coexistence of two different asymptotic
states; which one is reached by the system depends on the initial conditions. In region II both a stable incoherent state and a stable partially synchronized state
are possible, while in region VI one can have either a stable partially synchronized state or a stable standing wave state. The full dots of the dashed line denote
the points interested in the numerical simulation of the full oscillators system (see Section \ref{simdyn}).}
\label{fig_comp_struct}
\end{center}
\end{figure}

\subsection{Classification of bifurcations}
\label{classbif}

Referring to Fig. \ref{fig_comp_struct}, we give a brief description of the type of bifurcations associated to the passage, crossing the lines in the diagram,
from one kind of asymptotic state to another.
A very vivid explanation of the various types of bifurcation can be found in \cite{strogatzbook}.

Let us begin with the transition associated to the crossing of the circumference {\it `eabf'}. We have seen that at this crossing a stationary state
with positive $|r_1|$ bifurcates from the always existing incoherent stationary state with $r_1=0$. In the section {\it`ea'} of the circumference
the bifurcating positive $|r_1|$ state exists only inside the circumference, while in the section {\it `abf'} it exists only outside the circumference.
This is the characteristic of a pitchfork bifurcation, with one of the stationary states existing only on one side of the transition. Actually,
such bifurcation is in general associated with dynamical systems with symmetry, with two symmetric bifurcating states; in this case, $|r_1|$ is defined
to be nonnegative, and the symmetric state does not exist. In the supercritical case (section {\it `ea'}) the bifurcating solution is stable, while in the
subcritical case (section {\it `abf'}), the bifurcating solution is unstable. In section {\it `bf'} the incoherent state is unstable
on both sides of this transition (contrary to the canonical case of a pitchfork bifurcation); this is because it looses stability before, but this does not
change the nature of this bifurcation.

We now consider the straight line {\it `bcg'}. Crossing this line from above, the incoherent state looses stability, with the creation of a stable limit cycle, the
standing wave state. This is the typical case of a supercritical Hopf bifurcation. In section {\it `bc'} of the line a stable partially synchronized state with finite
$|r_1|$ already exists at the transition, together with the incoherent state, while in section {\it `cg'} the only stable state before the transition
is the incoherent one.

Crossing the line {\it `achd'} from the right, we have the appearance of two stable partially synchronized states with finite $|r_1|$, one
stable and the other unstable. We consider separately the section {\it `ach'} and the section {\it `hd'}. In the former case we have a saddle-node
bifurcation; the other stable state, i.e., the incoherent $r_1=0$ state in section {\it `ac'}, or the stable standing wave state in section {\it `ch'},
does not take part in the bifurcation. On the orther hand, in section {\it `hd'}, crossing the line from the right we have the disappearance of a stable
limit cycle and the appearance of the two partially synchronized states, one stable and the other unstable. Here we have a saddle-node infinite period
bifurcation, in which the period of the limit cycle tends to infinite at the bifurcation, i.e., it develops a fixed point, that after the transition
splits in the two partially synchronized states, the node and the saddle.

Finally, we consider the line {\it `bh'}. Crossing this line from the right, the limit cycle, i.e., the standing wave state, disappears. This is a
homoclinic bifurcation, in which the limit cycle, at the transition, reaches the saddle at $r_1=0$, with the appearance of a homoclinic orbit. After the
transition this orbit disappears, leaving the saddle at $r_1=0$. The partially synchronized state exists throught the transition and does not take part in it. 

\section{Larger values of $D$}
\label{largerd}

Having completed the analysis of the phase diagram for $D=0.5$, we now study how the structure of the diagram changes when $D$ is varied. But first let us
make a comparison with the phase diagram for $D=0$, analyzed in \cite{martens2009}. As remarked above, the Ott-Antonsen ansatz, applicable when $D=0$,
allows to have a closed four-dimensional dynamical system, reduced to two dimensions with further assumptions, something which is not possible for positive $D$.
In spite of this, we find that the phase diagrams for $D=0$ and $D=0.5$ are qualitatively the same, and also the type of bifurcations are the
same\footnote{We noted above that in some places, e.g., when dealing with the periodic solutions corresponding to the standing wave states, we had to dispense
with full mathematical rigour, due to the large dimensionality of our system, and rely to the numerical study of the system of
Eqs. (\ref{eqrn1h}) and (\ref{eqrn2h}).}. We note in particular that the circumference {\it `eabf'} and the line {\it `bcg'} are linearly shifted downward with
$D$, as one can see from expression (\ref{synchtrans}) defining the circumference and expression (\ref{eigen}) giving the stability thresholds for the incoherent
state: the center of the circumference is placed in $(\omega_0,\Delta)= (0,1-D)$, and the real part of the eigenvalue determining stability for $\omega_0>1$
is $1-D-\Delta$. On the other hand, the other lines of the phase diagram, i.e., {\it `achd'} and {\it `bh'}, do not share this exact property. For example,
while points {\it `a'} and {\it `h'} have coordinates $(\sqrt{3}/2,3/2)$ and about $(1.359,0.748)$, respectively, for $D=0$ \cite{martens2009}, their coordinates
for $D=0.5$ are $(0.905,0.925)$ and $(1.275,0.269)$, respectively; then, the coordinates for $D=0.5$ are not obtained by those for $D=0$ with a $(0,-0.5)$
shift, but the difference is not large.

As a consequence of the above considerations, we expect that by increasing $D$ the structure of the phase diagram will shift downward in the $(\omega_0,\Delta)$
plane. The first qualitative change will occur when point {\it `h'} reaches the $\omega_0$ axis. We have found that this occurs for $D \approx 0.765$. Then,
in Fig. \ref{fig_struct_d09} we plot the phase diagram for $D=0.9$.
\begin{figure}[h]
\begin{center}
\includegraphics[scale=0.5,trim= 0cm 3.0cm 0cm 0cm,clip]{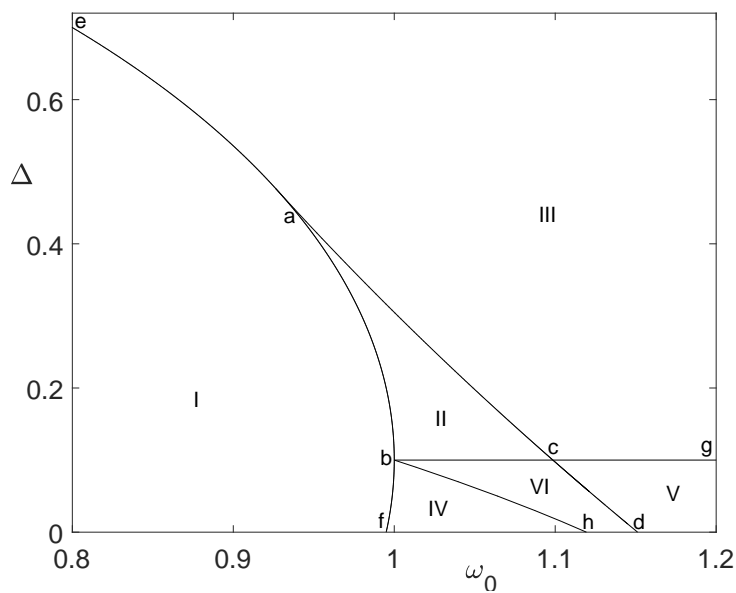}
\caption{The interesting region of the phase diagram for $D=0.9$. The meaning of the lowercase letters and roman numbers, is the same as in the diagram
for $D=0.5$ Also the stationary states or standing wave states are the same in the respective zone. The difference is that the line {\it `bh'} of the homoclinic
bifurcation reaches the $\omega_0$ axis before reaching the line {\it `acd'} of the saddle-node bifurcation.}
\label{fig_struct_d09}
\end{center}
\end{figure}
Since by now the overall structure of the diagram is clear, we plot directly a zoom of the interesting region, that includes all the transition lines.
The main difference with respect to the phase diagram for $D=0.5$ is that the line {\it `bh'} reaches the $\omega_0$ axis before the reaching the
line {\it `acd'} of the saddle-node transition. As a consequence, the diagram does not present any more the section {\it `hd'} of the transition line
that is found in Fig. \ref{fig_comp_struct}, and the corresponding saddle-node infinite period bifurcation. Thus, it is no more possible to have
a direct transition from region V to region IV, but only passing through region VI: the partially synchronized state will always appear before
the standing wave state disappears.

By increasing $D$ beyond $D=1$ the line {\it `bcg'} of the Hopf bifurcation disappears, together with regions IV, V and VI, and the standing wave states.
In Fig. \ref{fig_struct_d1p2} we plot the phase diagram for $D=1.2$. 
\begin{figure}[h]
\begin{center}
\includegraphics[scale=0.5,trim= 0cm 3.0cm 0cm 0cm,clip]{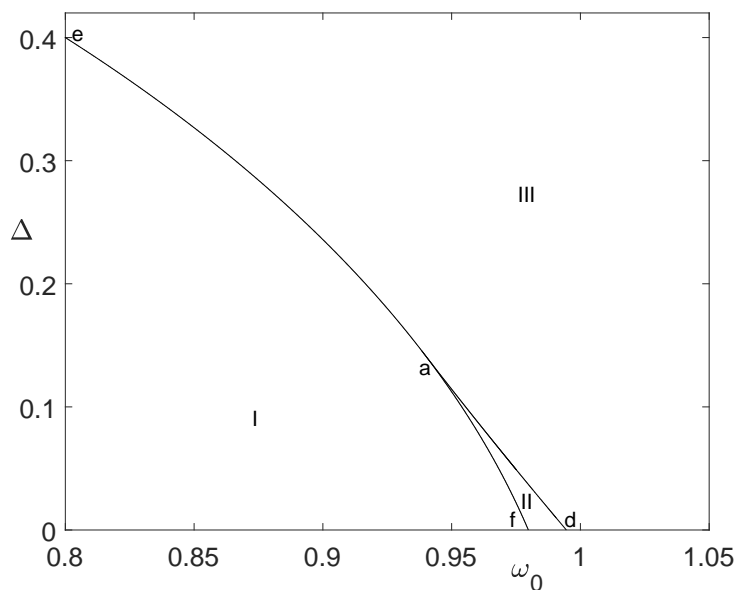}
\caption{The interesting region of the phase diagram for $D=1.2$. Only regions I, II and III are left, while the standing wave states, the Hopf bifurcation
and the homoclinic bifurcation have disappeared.}
\label{fig_struct_d1p2}
\end{center}
\end{figure} 
With respect to the phase diagram for $D=0.9$, regions IV, V and VI have disappeared.
There are no more limit cycles corresponding to standing wave states, and the Hopf bifurcation and the homoclinic bifurcation do not occur any more. Only in
region II we can now find the coexistence of two stationary states, the incoherent state and a partially synchronized state.

It is possible to find analytically up to which value of $D$ the point {\it `a'} exists. This threshold value is the one for which
Eq. (\ref{synchtrans}) and Eq. (\ref{synchsub}) considered as an equality have $\Delta=0$ as solution. This occurs for $D=4/3$. Beyond this value the only
feature of the phase diagram is the arc of circumference with radius $1$ and center in $(0,1-D)$ lying in the first quadrant. Clearly, we must also
have $D<2$, otherwise there is no such arc in the first quadrant. In Fig. \ref{fig_struct_d1p6} we plot the phase diagram for $D=1.6$. 
\begin{figure}[h]
\begin{center}
\includegraphics[scale=0.5,trim= 0cm 3.0cm 0cm 0cm,clip]{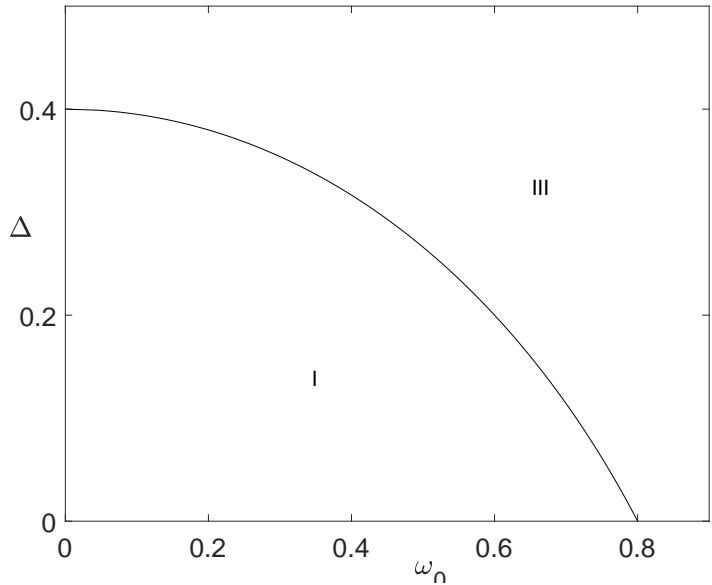}
\caption{The phase diagram for $D=1.6$. Only regions I and III, with a supercritical bifurcation between the incoherent state and the partially synchronized state,
are left. This is the same simple situation that occurs, with or without noise $D$, with a unimodal frequency distribution.}
\label{fig_struct_d1p6}
\end{center}
\end{figure} 
Now there are only regions I and III, where, respectively, we have the incoherent stationary state and the partially synchronized stationary state. Crossing
the transition line there is a supercritical bifurcation, corresponding to the transition between these two states. This is what is found in the case of a
symmetrical unimodal frequency distribution, independently from the value of $D$. Physically, this can be understood from the fact that a large noise tends
to mask the valley between the two peaks of the bimodal frequency distribution.

For $D>2$ no transition lines are any more present, and the only stationary state is the incoherent one.

\section{Numerical simulation of the dynamics}
\label{simdyn}

We have performed numerical simulations of the Langevin equations of the full system of coupled oscillators, Eqs. (\ref{langeq}),
that have shown that the dynamics of the order parameters is well
represented by the system of equations (\ref{eqrn1h}) and (\ref{eqrn2h}) truncated at a relatively small value of $n$. We will comment on this in the Discussion,
while in this section we focus
on the observation of hysteresis. In fact, the presence of regions, in the parameter space, where different asymptotic states can exist, with the system choosing
one of them depending on the initial conditions, makes it possible the existence of hysteresis.

In this section, since we have to refer to the original system parameters $\omega_0$, $\Delta$, $D$ and $K$, we reintroduce the use of the hat for the reduced variables,
that have been used throughout the paper, in particular in the plots of the phase diagram.

We have performed a simulation of Eqs. (\ref{langeq}) with $100000$ oscillators, with frequencies distributed according to
Eq. (\ref{lorendist}) with $\omega_0=2.9696$ and $\Delta=1$. Our purpose has been to simulate a dynamics where at predetermined times the coupling
$K$ is changed. In particular, we have started with a low value of $K$, for which the system is expected to reach an incoherent stationary
state\footnote{Obviously for the full system stationarity refers to the value of the order parameter, not to the dynamical degrees of freedom.},
and then we have increased $K$ several times at predetermined times; after reaching a maximum value of $K$, we have reversed the process, going back
to the same values of $K$ up to the initial value. The points of the dashed line in the lower panel of Fig. \ref{fig_comp_struct} refer to this simulation. We
have started the simulation with the system in the rightmost point of that line, denoted by a full dot, that belongs to region III; after a given time we have
changed $K$ so that to stay in the point of the dashed line inside region V; then another change of $K$ has taken the system in the point of the dashed line inside
region VI; and so on, we have then visited the point in region IV and the one in region I corresponding to the leftmost point of the dashed line. Then, we have
reversed the changes of $K$, going back to the point in region III.

The phase diagram of Fig. \ref{fig_comp_struct} is at constant $\widehat{D}=0.5$, therefore to simulate a system with given system parameters $\omega_0$ and $\Delta$,
but with constant reduced noise $\widehat{D}$, at each variation of $K$ we had to change correspondingly the value of $D$, since $\widehat{D}=\frac{4}{K}D$. The
analogous relations for the other parameters, i.e. $\widehat{\omega_0}=\frac{4}{K}\omega_0$ and $\widehat{\Delta}=\frac{4}{K}\Delta$, give the variation
of $\widehat{\omega_0}$ and of $\widehat{\Delta}$ when $K$ is varied, and then determine how one moves on the phase diagram.

In Fig. \ref{fig_hyster} we plot the behaviour of the order parameter $|r_1|$ during the simulation.
\begin{figure}[h]
\begin{center}
\includegraphics[scale=0.5,trim= 0cm 3.0cm 0cm 0cm,clip]{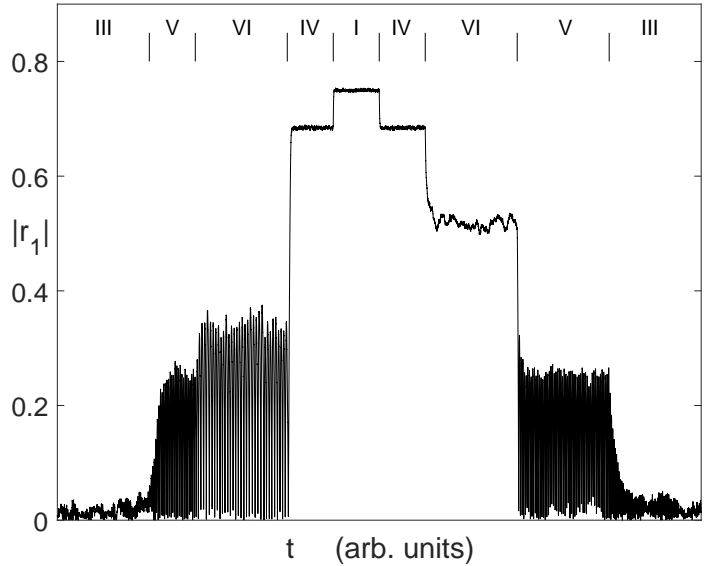}
\caption{The order paramete $|r_1|$ as a function of time in a numerical simulation of Eqs. (\ref{langeq}), with $100000$ oscillators. The small vertical bars
at the top of the plot denote the times in which the coupling $K$ has been changed. The roman numbers show the region of the phase diagram in which the system
parameters are located during that lapse of time; more precisely, the parameters of the system are those corresponding to the respective point of the dashed line
in the lower panel of Fig. \ref{fig_comp_struct}. The time lapses are not all equal.}
\label{fig_hyster}
\end{center}
\end{figure} 
At the beginning, when the system is in region III, the order parameter is almost $0$; of course it cannot be exactly $0$, due to finite size effects, that are
responsible of the fluctuations also in the following phases of the run. When the coupling $K$ is increased so that the system is inside region V, we see that
the dynamics enters a phase with an almost periodic variation of the order parameter. Again, the imperfect periodicity is caused by finite size effects; however,
it is nice to see that the minimum value of $|r_1|$ during the oscillations is practically $0$, as predicted by the reduced system of equations.
A further increas of $K$, that brings the system in region VI, has the effect of enhancing the amplitude of the periodic variation of $|r_1|$. In this region,
we had found the coexistence, together with the standing wave state, of a partially synchronized state; this is observed later in the simulation, marking the
hysteresis of the dynamics. A further increase of $K$ brings the system in region IV, and then in region I; this corresponds to the order parameter
staying in a stationary value, corresponding to a partially synchronized state. When the reversed process is begun, going back to region IV, the previous
value of $|r_1|$ is obtained, as shown in Fig. \ref{fig_hyster}. However, when we now bring back the system in region VI, it settles to the partially synchronized
state, although with fluctuations of $|r_1|$ somewhat larger than in regions I and IV; this can also be due to the fact that region VI is rather narrow, so that
the values of the parameters are not very far from those at the boundary of the region. Taken back to region V, the system goes back to the only existing
asymptotic state, the standing wave state, while brought back finally in region III, $|r_1|$ goes back to $0$.

Summarizing, in the above dynamics the system is found twice in region VI, where different asymptotic states exist. The fact that in the first passage in the region
the system settles in the periodic state, while in the second passage it goes to the stationary partially synchronized state, proves the existence of hysteresis loops.

\section{Discussion and conclusions}
\label{discuss}

The peculiarity of the frequency distribution used in this work, Eq. (\ref{lorendist}), is the fact that, when analytically continued to the complex
$\omega$ plane, it has few poles. Of course it is not difficult to envisage other frequency distributions with the same property. We have seen that, even without
the possibility of using the Ott-Antonsen ansatz, this gives the possibility to study directly the dynamics of the order parameter of the system, the main
variable related to the synchronization transition. In a noiseless system the ansatz and the mentioned property of the frequency distribution allow, together,
to reduce the study to a low dimensional system of equations, while in a noisy system the system of equations is still infinite dimensional, in principle. However,
the dynamical variables of this system, Eqs. (\ref{eqrn1h}) and (\ref{eqrn2h}), are the order parameters $r_n^{(1)}$ and $r_n^{(2)}$, and we have argued that
at increasing $n$ these variables approach rapidly zero, since a finite $r_n$ for large $n$ would require a distribution function with very large fluctuations
(of course, one could study the Fokker-Planck equation (\ref{fpstart}) with such a distribution as initial condition, but in this case we expect, on physical grounds,
that the fluctuations will smooth out rapidly). This allows to truncate the system at a reasonable small value (we have chosen $M=50$ as the largest value of $n$),
without spoiling the resulting dynamics.

We have found that at small values of the reduced noise parameter $D$ the phase diagram of the system is qualitatively similar to that of the noiseless
$D=0$ case. The main quantitative difference is an approximate overall downward shift of the transition lines, the overall shift quantified by the noise $D$ itself.
This downward shift is exact for the transition lines corresponding to the pithfork bifurcation and the Hopf bifurcation, while it is only approximate for the
transition lines corresponding to the saddle-node bifurcation and the homoclinic bifurcation (for example, point {\it `a'} reaches the $\Delta =0$ axis for $D=4/3$,
but at $D=0$ it is found at $\Delta=3/2$ \cite{martens2009}).

The progressive simplification of the phase diagram at increasing values of $D$ can be understood on physical grounds. We know that for a symmetric unimodal
frequency distribution the only transition is the one between the incoherent state and the partially synchronized state, and that as soon as the latter exists,
on one side of the (supercritical) transition, the former looses stability. This picture is independent on the level of noise, that determines only the location of
the transition. A large noise can be physically interpreted as un uncertainty in the proper frequency of each oscillator, and then as a blurring of the frequency
distribution (as we have remarked in the Introduction, one main motivation for the introduction of noise in a system of coupled driven oscillators is the possibility
to represent in this way the uncertainty and the fluctuations of the proper frequencies). This process tends to decrease the depth of the valley between the two
peaks of a bimodal distribution, until the valley is completely washed out at large enough noise, and the syastem behaves as if it had a unimodal distribution.

A natural question that arises is what happens if the frequency distribution does not have the property of having just few poles when analitycally prolongued
in the complex plane. In that case, the restriction of the dynamics to that of the order parameters, and practically to the first several $r_n$, is not possible,
and one should analyze the full Fokker-Planck equation (\ref{fpstart}). Again we can try to resort to a physical argument. A frequency distribution made of,
e.g., the sum of two Gaussians centered in $\pm \omega_0$, has an essential singularity at the point at infinity in the complex plane, and a study analogous
to that in this work cannot be performed. However, it is possible to approximate numerically such a disribution with one decaying algebraically; the approximation
would fail only at large frequencies, that will be the proper frequencies of few outlier oscillators. One can argue that these two systems should behave
very similarly, presenting the same types of stationary or periodic asymptotic state, and the same types of transitions between them, with just small
differences in the location of the transitions. The consequence of this argument is that, apart from numerical details, one could study the behaviour of a
general system by trying to approximate as close as possible the frequency distribution with an algebraic one. Under this perspective, it is not by chance that
for the noiseless system the numerical results for a sum of two Gaussians are close to those of the sum of two Lorentzians \cite{martens2009}.

Adopting this point of view, it would be interesting to perform an analysis like the one presented in this work for more general frequency
distributions $g(\omega)$, although still with the property of having few poles in the analytical continuation, and more general forms of the interaction between
the oscillators. This could allow the study of the complete phase diagram, that could be even richer than the one occuring for symmetrical bimodal distributions.
For example, the interaction given in Eqs. (\ref{langeq}), i.e., the interaction used in the Kuramoto model, is the simplest one if one considers the Fourier
expansion of a generic interaction $h(\theta_i - \theta_j)$. This will have, in general, all the Fourier terms proportional to $\sin [k(\theta_i -\theta_j)]$
for each integer $k$ (the terms proportional to the cosines are excluded if we want interactions derived from a potential). It has been found that
noiseless $D=0$ systems with this generic interaction have an order parameters that scales differently, with respect to the Kuramoto model, near the onset of the
synchronization transition \cite{daido1,daido2}. Extension to the noisy case with generic interactions has shown that in this scaling behavior there is a crossover,
since the scaling tends to go back to the Kuramoto result when the noise strength increases \cite{craw1,craw2}. The study of the full phase diagram, at various
noise strengths, could be very rewarding. On the other hand, restricting to the simple sine interaction, but considering nonsymmetrical frequency distributions,
should give rise to new asymptotic states, like travelling waves. It would be equally interesting to study the effect of noise in this case, to see, e.g., 
what would be the effect of the blurring of the frequency distribution at large noise.

\ack
The author acknowledges financial support from INFN (Istituto Nazionale di Fisica Nucleare) through the projects DYNSYSMATH and ENESMA.

\end{document}